# Lower Bound of Thermal Conductivity in Graphene

Yan Gao[1,2,3,4†], Yafei Ding[3,4,5†], Hongbin Xu[2,3,6†], Zhaoxuan Tang[3,7], Fuqing Duan[5,3,4], Wenmei Luo[2,3,8], Zhensheng Zhang[4], Xiaohang Zhang[9*], Guimei Zhu[3,8*], Yahui Xue[1,3*], Baowen Li[2,3,5,10*]

[1]*Department of Mechanics and Aerospace Engineering, Southern University of Science and Technology, Shenzhen, 518055, China.*

[2]*Department of Physics, Southern University of Science and Technology, Shenzhen, 518055, China.*
[3]*Shenzhen Key Lab for Phononics and Intelligent Thermal Materials, Shenzhen, 518055, China.*
[4]*International Quantum Academy, Shenzhen, 518000, China.*
[5]*Department of Materials Science and Engineering, Southern University of Science and Technology, Shenzhen, 518055, China.*

[6]*College of Semiconductors (National Graduate College for Engineers), Southern University of Science and Technology, Shenzhen, 518055, China.*

[7]*School of Power and Mechanical Engineering, Wuhan University, Wuhan, 430072, China.*
[8]*School of Microelectronics, Southern University of Science and Technology, Shenzhen, 518055, China.*
[9]*The Institute for Brain Research, Advanced Interfaces and Neurotechnologies, Shenzhen Medical Academy of Research and Translation, Shenzhen, 518100, China.*

[10]*State Key Lab for Quantum Functional Materials, Shenzhen, 518055, China.*
[†]*These authors contributed equally to this work.*
[*]*e-mail: zhangxiaohang@smart.org.cn, zhugm@sustech.edu.cn, xueyh@sustech.edu.cn, libw@sustech.edu.cn*

**Abstract:** The intrinsic thermal conductivity ($\kappa$) of graphene remains highly controversial because of physical and metrological complexities. Although the thermal bridge method enables reliable measurements, sample fabrication remains challenging. Here, we develop a highly efficient patterning and transfer technique to systematically fabricate and characterize suspended bilayer graphene (BLG) thermal bridges with lengths of 1–30 μm. Suspended BLG with unavoidable polymer residues exhibits length-independent thermal conductivity of ~570 W/(m·K) at room temperature, remaining nearly constant from 250 to 350 K and robust to rigorous surface cleaning and amorphous silicon nitride substrates. Homogeneous non-equilibrium molecular dynamics (HNEMD) simulations show that amorphous carbon coverage of only ~0.01 monolayer is sufficient to suppress out-of-plane flexural acoustic (ZA) phonons. These findings challenge the prevailing view that rigid substrates suppress ZA phonons more strongly than surface contamination and identify a robust lower bound of κ governed by amorphous surface scattering. This lower bound reflects heat conduction in graphene under realistic device conditions and provides a reference for fundamental phonon transport studies and the integration of graphene into thermal-management applications in electronic devices and chips.

Advanced thermal management of microelectronic chips has become indispensable due to the escalating integration and power density, alongside the rapid proliferation of artificial intelligence (AI)[1, 2].

Consequently, understanding the physical mechanisms of micro- and nanoscale phonons transport and developing high-efficiency thermal management materials are of great relevant and imperative[3, 4], because phonons are the primary thermal energy carriers in semiconductors and the relevant devices.

Graphene has been widely heralded as a premier candidate for next-generation thermal management ever since its ultrahigh thermal conductivity was first unveiled via the Raman optothermal method[5]. However, despite extensive research efforts, the value of its intrinsic thermal conductivity remains a longstanding contentious issue: reported experimental values of thermal conductivity ($\kappa$) exhibit a drastic discrepancy, ranging from 200–300 W/(m·K) to 5000 W/(m·K) [5-18]. This immense divergence fundamentally stems from compounded physical and metrological complexities[19].

Physically, phonon transport is intricately modulated by size effects[8, 14, 20], surface contamination[11, 16, 21], and substrate coupling[13]. Metrologically, existing characterization techniques suffer from inherent limitations. Although the Raman optothermal method simplifies sample preparation, it is plagued by substantial measurement uncertainties[5, 22]. In contrast, while the suspended thermal bridge method is widely acknowledged as the most rigorous technique for probing the in-plane $\kappa$ of two-dimensional (2D) materials, its broad application is severely bottlenecked by the extreme mechanical challenges of sample fabrication[21, 23, 24]. Furthermore, the inability to accurately decouple contact thermal resistance ($R_c$) compromises measurement fidelity[23, 25-27], particularly for highly conductive materials like graphene. Together, these factors precipitate a chaotic divergence in experimental data, severely impeding both the pure fundamental research and the practical applications of graphene-based thermal management.

In this work, we present a highly efficient patterning and transfer technique to systematically fabricate suspended bilayer graphene (BLG) thermal bridge samples with lengths ranging from 1 to 30 μm. This represents the most extensive array of suspended graphene samples characterized via the thermal bridge method to date, enabling the rigorous decoupling of $R_c$ utilizing the intercept method. Our systematic measurements reveal that, at the micrometer scale, suspended BLG with unavoidable polymer residues exhibits a length independent thermal conductivity, yielding a saturated value of ~570 W/(m·K). Remarkably, this $\kappa$ value remains unaltered despite rigorous surface cleaning treatments, variations in surface contaminants, or the introduction of amorphous silicon nitride substrates. Homogeneous non-equilibrium molecular dynamics (HNEMD) simulations further elucidate this phenomenon, demonstrating that a minimal surface coverage of amorphous carbon of only ~0.01 ML is sufficient to saturate the scattering of out-of-plane flexural acoustic (ZA) phonons, resulting in a drastic ~70% reduction in $\kappa$. Ultimately, these findings determine a robust lower bound for BLG governed by amorphous surface scattering. Considering the unavoidable surface contamination of graphene in practical applications, this lower bound serves as an essential experimental baseline for thermal management design and material development.

## Results

High-quality graphite flakes were mechanically exfoliated from natural graphite onto a silicon substrate with $SiO_2$ layer. Subsequently, we patterned and transferred graphene utilizing a method developed specifically for suspended two-dimensional (2D) materials (Figure 1a). Mechanically exfoliated graphene was initially protected by a poly (methyl methacrylate) (PMMA) mask fabricated via electron-beam lithography (EBL) (Figure 1b). Subsequently, reactive ion etching (RIE) was employed to remove the unprotected regions of the graphene, thereby achieving the desired patterning (Figure 1c). A layer of PMMA was then spin-coated onto the patterned graphene to serve as a protective coating and temporary mechanical support. Unless otherwise specified, PMMA 950K is utilized for the transfer of graphene samples. Following this, a polydimethylsiloxane (PDMS) stamp was attached to the PMMA surface. By employing deionized water intercalation at the PMMA/$SiO_2$ interface, the PDMS/PMMA/graphene composite structure was gently delaminated from the silicon oxide substrate[28]. Utilizing a 2D material transfer platform, the graphene was precisely aligned with the target location on the thermal bridge device, and the PMMA/graphene heterostructure was released onto the surface of the thermal bridge (Figure 1d). The device was then sequentially immersed in acetone and isopropyl alcohol (IPA) to dissolve the PMMA layer. Finally, to prevent structural damage to the graphene induced by capillary forces during the evaporation process, a critical point dryer (CPD) was utilized to dry the samples. More experimental details can be found in the Methods section.

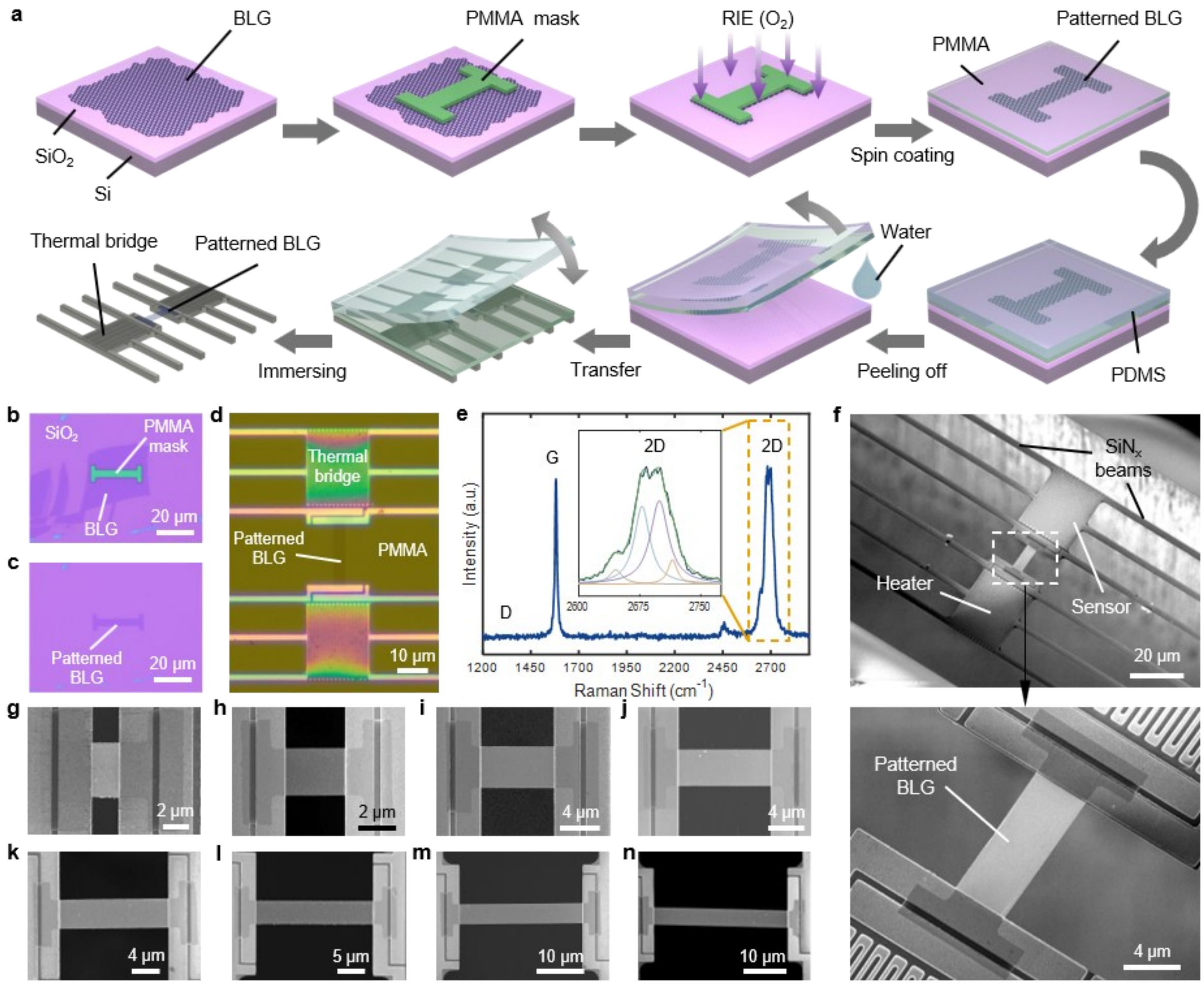

**Figure 1. Sample preparation and characterization.** **a**, Schematic illustration of the graphene patterning and transfer process. **b**, Optical micrograph of the BLG covered with a PMMA mask. **c**, Optical micrograph of the patterned BLG. **d**, Optical micrograph of the PMMA-assisted graphene transfer onto the thermal bridge. **e**, Raman spectrum of the fabricated BLG sample and the corresponding Lorentzian fitting curves of the 2D band. **f**, SEM image of the BLG on the thermal bridge device. **g**–**n**, SEM images of partial BLG samples with varying lengths (2, 5, 7, 10, 15, 20, 25, and 30 μm). The width of all samples is fixed at 4 μm.

Raman spectroscopy (Figure 1e) demonstrates that the D band of the graphene sample is nearly negligible, indicating a highly pristine crystal lattice with minimal defects (Figure S1). The 2D band is well-fitted by four distinct Lorentzian components, indicating that the graphene is bilayer[29, 30]. Although our current fabrication capabilities enable the reliable preparation of thermal bridge devices ranging from monolayer graphene to thin graphite flakes, samples with a thickness of bilayer and above exhibit a significantly higher fabrication yield. Therefore, to ensure a statistically sufficient number of samples for our systematic investigation, BLG was selected as the primary subject of this study. Meanwhile, the width of the samples was kept constant at 4 μm. This dimension was chosen based on prior research indicating that the thermal conductivity of graphene reaches a converged value at this scale, exhibiting no further width dependence[14, 31]. As illustrated by the scanning electron microscopy (SEM) image in Figure 1f-n, the dried graphene sample exhibits a well-defined geometric morphology.

The in-plane thermal conductivity of the BLG was measured utilizing a micro-thermal bridge device. As illustrated in Figure 2a, the device is composed of two fully suspended silicon nitride ($SiN_x$) platforms, which are respectively connected to the substrate by six $SiN_x$ beams. Platinum (Pt) serpentine electrodes are patterned on the $SiN_x$ surfaces to be utilized for heating and temperature sensing. The two $SiN_x$ platforms are bridged by the graphene sample, whereby a completely suspended configuration is achieved. During the measurement, the serpentine electrode on one side is designated as the heater, while the electrode on the opposite side serves as the sensor (further details are provided in the Methods). The total thermal resistance ($R_{total}$) of the sample can be calculated by $R_{total} = (\Delta T_h^2 - \Delta T_s^2) / (Q\Delta T_s)$[23, 24], where $\Delta T_h$ and $\Delta T_s$ denote the temperature rises of the Heater and Sensor during the measurement, respectively. The total Joule heat generated on the suspended platform by the DC current is represented by $Q$, which is determined by the Heater's electrical resistance and the applied DC magnitude.

Enabled by the highly efficient patterning and transfer technique, we fabricated thermal bridge samples with lengths spanning from 1 to 30 μm to systematically study the length dependence and the $\kappa$ of BLG. The measurements of the total thermal resistance ($R_{total}$) are illustrated in Figures 2b–d (additional temperature points data are shown in Figure S2). These plots present the thermal resistance data of about 30 BLG samples prepared via PMMA-assisted transfer, featuring a constant width of 4 μm but varying lengths. Except for a slightly inferior linear fit at the low temperature of 30 K caused by larger data fluctuations, a linear increase of $R_{total}$ with the sample length can be observed at higher temperatures. This linear dependence demonstrates that no obvious size effects are exhibited at the micrometer scale, implying that the intrinsic thermal conductivity of the sample remains constant. Consequently, the derivative of $R_{total}$ with respect to length is given by $dR_{total} / dL = 1/(\kappa W h)$, where $L$, $W$, and $h$ are the length, width, and thickness of the suspended portion of the sample, respectively. The $\kappa$ at a given temperature can be extracted from the slope of the linear fit of the length-resistance data, while the $R_c$ at that temperature is represented by the intercept (results are shown in Figure 2f).

The temperature dependence of the BLG thermal conductivity is plotted in Figure 2e. At 300 K, the thermal conductivity of the BLG is determined to be 572.7±10.3 W/(m·K) (Details regarding the error analysis can be found in Supporting Information), which is substantially lower than previously reported theoretical predictions of approximately 2000 W/(m·K)[17, 32, 33]. Notably, a stable value of around 570 W/(m·K) is temperature independent within the temperature range of 250 K to 350 K. Furthermore, the relationship between $R_c$ and temperature is shown in Figure 2f, yielding $R_c$ of $2.05 \pm 0.13 \times 10^6$ K/W at 300 K. Taking a BLG sample with $L$ = 10 μm as an example, a total thermal resistance of $8.47 \pm 0.15 \times 10^6$ K/W at 300 K can be derived from the fitting result in Figure 2d. Here, ~24% of the $R_{total}$ is accounted for by $R_c$. It is highlighted by this proportion that a significant underestimation of the measured $\kappa$ would be caused if the $R_c$ were ignored. Moreover, since $R_c$ is independent of the sample length $L$, an increasingly severe measurement error is introduced for shorter samples when $R_c$ is neglected. It should be clarified that, by convention, all thermal resistances outside the suspended portion of the sample are collectively referred to as $R_c$. However, this portion of the thermal resistance is not solely contributed by the physical interface between the sample and the thermal bridge. Because a finite spatial distance is present between the Heater/Sensor temperature-measuring electrodes and the sample, an additional parasitic thermal resistance is introduced into our measurements by this spatial gap[9]. Therefore, the $R_c$ defined here should be regarded as a lumped parameter, which is jointly composed of a partial internal thermal resistance from the thermal bridge device itself alongside the true interfacial thermal resistances across various interfaces along the heat conduction pathway.

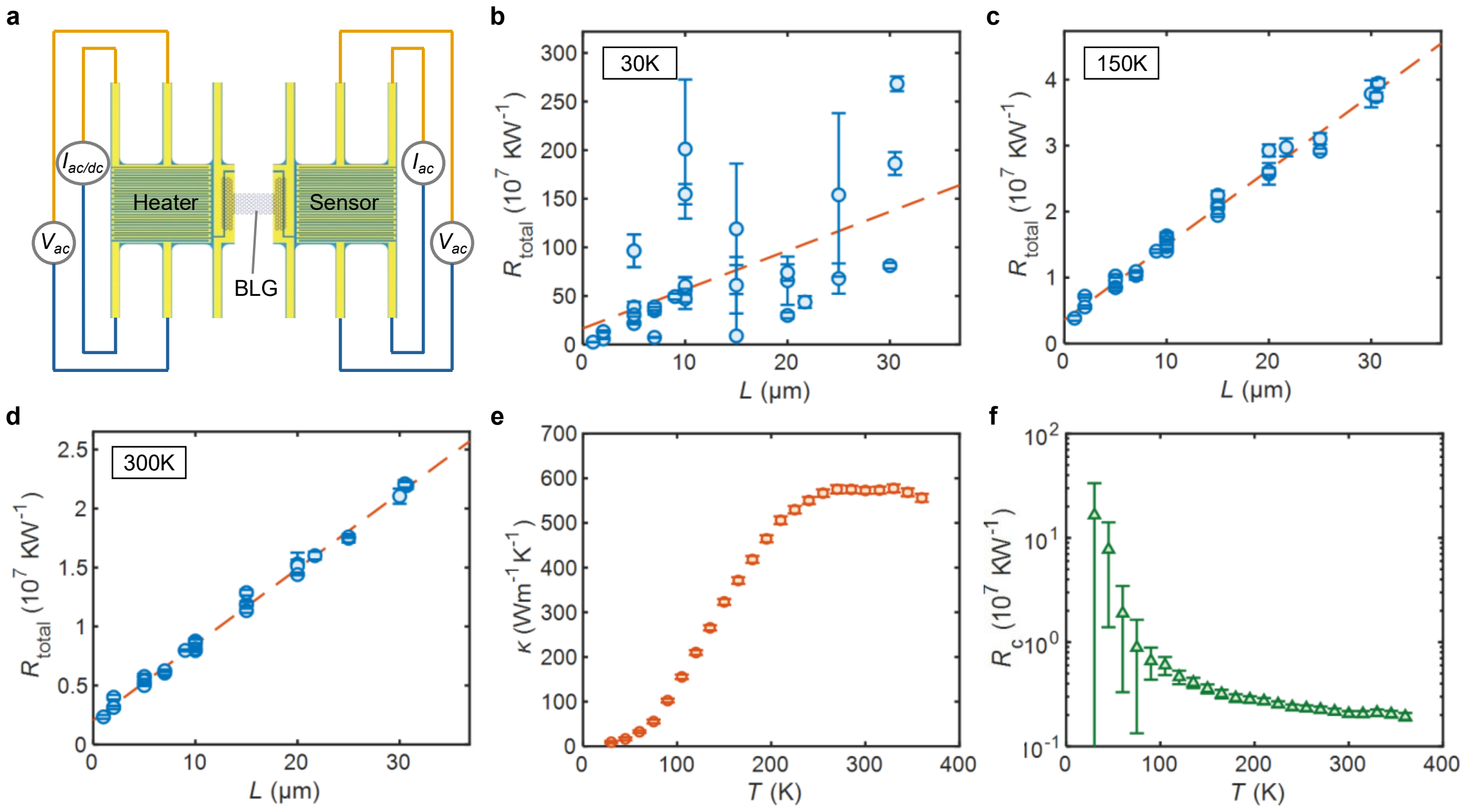

**Figure 2. Measurement setup and experimental results.** **a**, Schematic illustration of the micro-thermal bridge method utilized for measuring the in-plane thermal conductivity of the BLG. **b–d**, Total thermal resistance measured as a function of length for the PMMA-assisted transferred suspended BLG samples at 30 K, 150 K, and 300 K, respectively. A fixed width of 4 μm is maintained for all samples. Each data point corresponds to an individual sample, and the error bars represent the standard deviation derived from three repeated measurements. **e**, Temperature dependence of the BLG thermal conductivity, which is extracted and calculated from the slope of the linear fitting of the length-dependent thermal resistance measurements. **f**, Contact thermal resistance ($R_c$) vs temperature, which is obtained from the intercepts of the linear fits across different sample lengths. The error bars in **e** and **f** represent the standard errors propagated from the fitted slope and intercept of the linear total thermal resistance–length relationship.

The suppressed thermal conductivity observed in our BLG samples is likely ascribed to extrinsic phonon scattering induced by PMMA residues. To validate this hypothesis, we implemented several strategies to optimize the surface cleanliness of the samples. Multiple cleaning methods were employed to remove the PMMA residues from the graphene surface, including conventional thermal annealing in $H_2$/Ar or $CO_2$ atmospheres[34-36]. Additionally, laser-induced cleaning was performed in both vacuum and ambient environments to remove surface contaminants[37]. As revealed by SEM images, the most pronounced cleaning effect was achieved by laser irradiation in a vacuum environment (as shown in Figure 3a and 3b). Following a 300-second exposure to a 3 mW laser, a large-area improvement in surface cleanliness was observed, with the white PMMA residues in the irradiated region being almost completely eliminated. Raman spectra before and after cleaning are provided in Figure S3. Despite this visual improvement, no significant changes in the thermal conductance of the samples were detected before and after the cleaning treatments (Figure 3c).

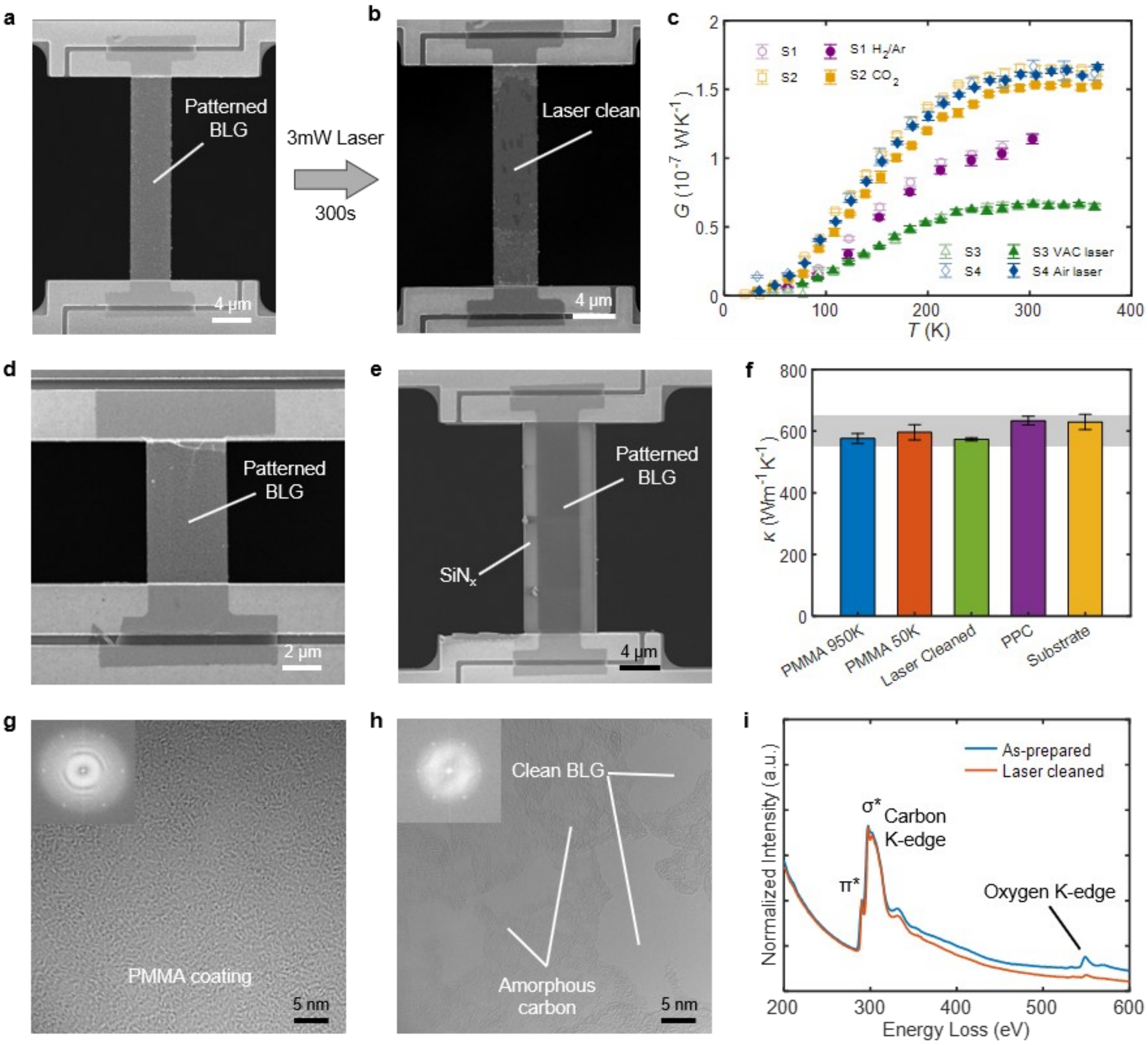

**Figure 3. Optimization of sample cleaning. a**, SEM image of a PMMA-assisted transferred BLG sample prior to laser cleaning. **b**, SEM image of the BLG sample following 300 s of 3 mW laser irradiation in a vacuum environment. **c**, Comparison of the thermal conductance ($G$) as a function of temperature for four individual samples before and after respective cleaning treatments: $H_2$/Ar annealing, $CO_2$ annealing, laser cleaning in vacuum, and laser cleaning in air. **d**, SEM image of a graphene sample transferred using poly(propylene carbonate) (PPC). **e**, SEM image of a BLG sample supported on an amorphous silicon nitride substrate. **f**, Influence of various surface loads/contaminants on the thermal conductivity of BLG at 300 K. The presented results are inherently corrected for $R_c$. The error bars in **c** and **f** represent the standard deviation derived from three repeated measurements. Transmission electron microscopy (TEM) images and corresponding fast Fourier transform (FFT) patterns of the BLG sample before **g** and after **h** laser cleaning. **i**, Electron energy loss spectroscopy (EELS) spectra of the BLG surface before and after laser cleaning. Following the cleaning process, the oxygen signal associated with PMMA residues is eliminated.

Furthermore, alternative transfer media, including low-molecular-weight PMMA 50K and polypropylene carbonate (PPC) (Figure 3d)[38], were utilized to replace the standard PMMA 950K in an attempt to further improve the initial sample cleanliness. For comparison, supported BLG samples on amorphous $SiN_x$ were also fabricated (as shown in Figure 3e) to investigate whether strong van der Waals interfacial coupling could further suppress the in-plane thermal conductivity of the BLG. The calculation method for the $\kappa$ of the supported BLG is detailed in the Supporting Information. The thermal conductivities of the BLG with different surface contaminants or substrate supports at 300 K are presented in Figure 3f. It can be observed that, although the specific materials causing surface phonon scattering differ, no obvious distinction in thermal conductivity was found among the samples. All measured values fall within a narrow range of 600±50 W/(m·K). This phenomenon implies that the scattering of acoustic phonons by surface contamination in our samples has already reached a saturation. Consequently, even when a rigid amorphous silicon nitride substrate was introduced, the thermal conductivity of the BLG could not be further reduced.

TEM was utilized to characterize the PMMA-assisted transferred BLG before and after laser cleaning, as shown in Figure 3g and 3h. Prior to cleaning, PMMA residues were observed to be discretely distributed across the graphene surface. Upon laser irradiation, a partial removal of amorphous contaminants was observed. Notably, electron energy loss spectroscopy (EELS, Figure 3i) reveals a near-complete eradication of the PMMA-associated oxygen signal after cleaning[39]. This suggests that the localized laser heating drives the dehydrogenation and deoxygenation of PMMA, ultimately carbonizing it into amorphous carbon. Crucially, while this treatment modifies the chemical makeup of the surface adsorbates, the resulting amorphous carbon layer persists as a highly dense source of extrinsic phonon scattering. This persistence elegantly elucidates the negligible change observed in the in-plane thermal conductivity before and after the cleaning procedure.

To elucidate the microscopic modulation mechanism of phonon heat transport in BLG by surface coverage $\theta$ (in ML), molecular dynamics (MD) simulations based on the neuroevolution potential (NEP) were performed for different surface coverages. To accurately account for the intrinsic phonon-scattering background of natural BLG, $^{13}C$ isotope doping at natural abundance (1.1%) is incorporated into the model. The carbon species with coverage of 0 ML, 0.001 ML, 0.005 ML, 0.01 ML, 0.04 ML, and 0.07 ML are adsorbed onto the BLG surface (Figure 4a), where the coverage is defined as the numerical ratio of carbon atoms on the surface to those in the top graphene layer. The optimized structure shows that, as coverage increases, the dispersed carbon single atoms gradually agglomerate into carbon chains and are anchored to the BLG surface via end-group adsorption. Considering the fitting error (Figure S5, S6) of NEP to the atomic force and heat flow, the corrected HNEMD calculation results show that the strong surface phonon scattering leads to the logarithmic decay of the thermal conductivity of BLG with the increase of $\theta$ between 0 ML and 0.01 ML (Figure 4b inset). More importantly, the thermal conductivity is stabilized around 585 W/(m·K) after 0.01 ML (Figure 4b). Both the surface adsorption models presented above and our previous computational work on substrate-supported graphene[40] yield saturated thermal conductivities that are highly consistent with current experimental data. The thermal conductivity is decomposed into contributions associated with out-of-plane and in-plane atomic vibrations, which are comparable in pristine BLG (Figure 4c). With increasing $\theta$, the out-of-plane contribution rapidly decreases to approximately 20% of the total thermal conductivity and remains near this level at coverages of 0.01 ML and above. This behavior is consistent with enhanced scattering of flexural phonons by surface carbon adsorbates. The more pronounced reduction in the out-of-plane contribution suggests that these vibrations are more sensitive to adsorption-induced structural perturbations than in-plane vibrations supported by the strong $sp^2$-bonded carbon network.

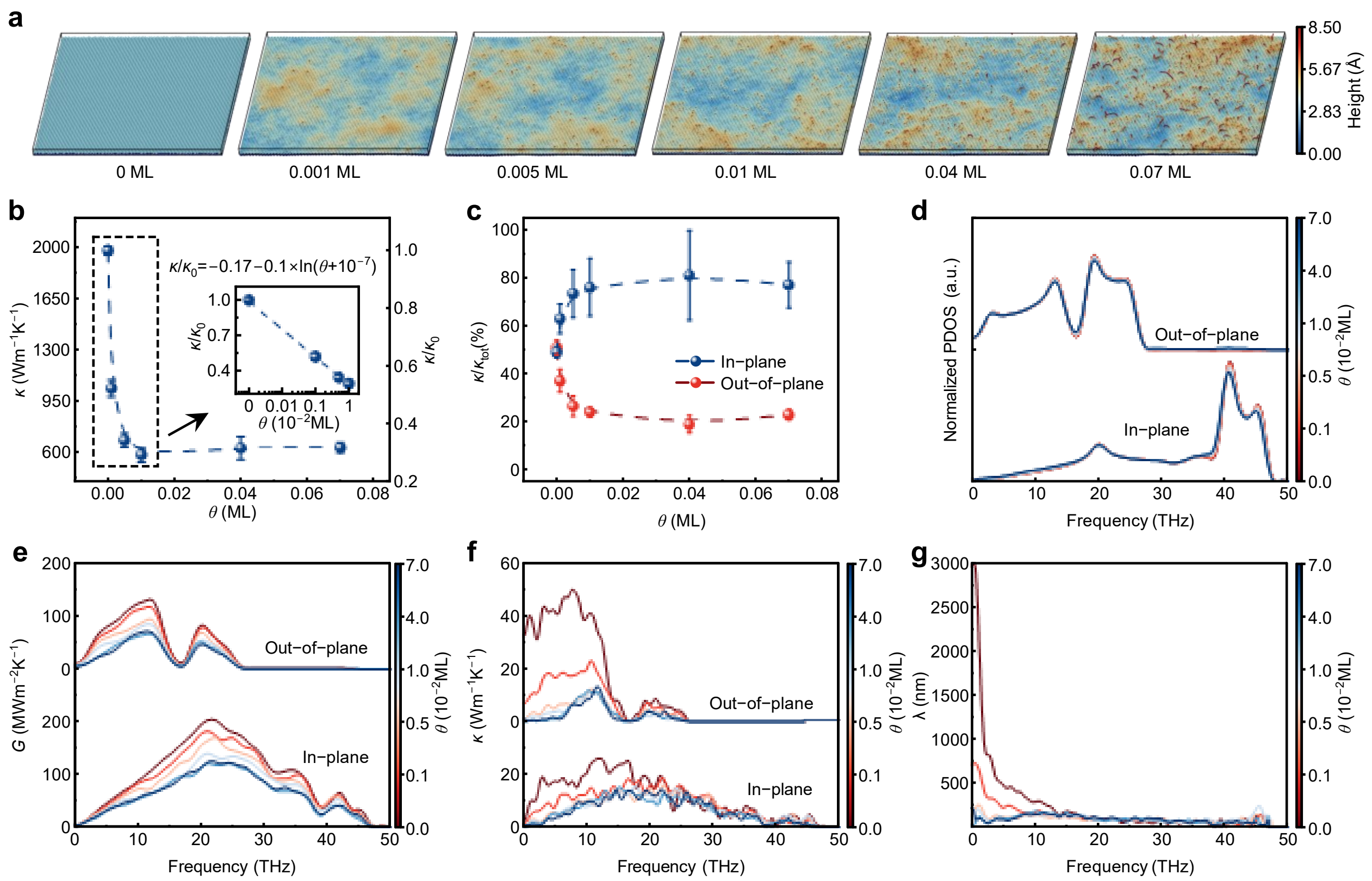

**Figure 4. MD simulation of the effect of surface coverage (*θ*) on phonon transmission in BLG. a**, Schematic of the BLG model with different $\theta$, the color mapping indicates the atomic height. **b**, Corrected thermal conductivity for different coverages, inset shows fitted curve for low coverages $\kappa/\kappa_0 = -0.17 - 0.1 \times \ln(\theta+10^{-7})$. Error bars show standard errors from four independent runs. **c**, In-plane and out-of-plane thermal conductivity ratios. **d-g**, Coverage-dependent spectral properties, color-coded by $\theta$: in-plane and out-of-plane (**d**) phonon density of states (PDOS), (**e**) ballistic thermal conductance, (**f**) diffusive thermal conductivity, and (**g**) spectral phonon mean free path.

To understand the physical mechanism of $\kappa$ reduction, we further calculated the phonon density of states (PDOS), as is shown in Figure 4d. It is found that the in-plane acoustic modes have a negligible effect on the PDOS. However, the two main out-of-plane peaks (13 THz and 19 THz) corresponding to the acoustic and optical modes (ZA and ZO, respectively) decrease monotonically with increasing $\theta$, indicating that surface species suppress the out-of-plane motion of graphene. Comparing the finite-length spectral ballistic thermal conductance with the infinite-length spectral diffusive thermal conductivity, the former intuitively shows the suppression of the full frequency range at the adsorbed surface (Figure 4e). At the same time, the main difference between the latter is concentrated in the low- and medium-frequency range (Figure 4f). These low-frequency phonons typically have relatively large group velocities or long relaxation times[41] and thus dominate the lattice heat conduction. As $\theta$ increases, the low-frequency spectral thermal conductivity decreases significantly, indicating that surface adsorbates can effectively scatter long-wavelength phonons. In addition, the stronger suppression of the out-of-plane spectral thermal conductivity further confirms that the surface species disrupt the transport of out-of-plane phonons. The spectrally resolved phonon mean free path further supports this interpretation (Figure 4g). In clean BLG, low-frequency phonons exhibit very long mean free paths, reaching the micron scale. With the introduction of surface species, the mean free path of these low-frequency phonons is significantly shortened. In contrast, high-frequency phonons have much shorter mean free paths and are therefore less affected by additional surface scattering. It indicates that the significant reduction in the total thermal conductivity arises mainly from the strong scattering of long-mean-free-path phonons by surface adsorbates. This result also explains why the size effect is not observed experimentally at the micrometer scale.

## Discussion

Despite the large variation in reported measurements of graphene thermal conductivity, the room-temperature value measured in this study is broadly consistent with those previously reported by Pettes et al.[11] for residue-bearing suspended bilayer graphene (BLG). The present study extends this experimental picture through length-dependent resistance measurements over 1–30 μm, together with comparisons involving different cleaning treatments, transfer polymers, and amorphous $SiN_x$ support. Conventionally, it is believed that rigid substrates suppress out-of-plane phonons in graphene more profoundly than surface contamination[6, 32, 42, 43]. However, our findings reshape this prevailing perception. Provided the graphene lattice structure remains intact, this value represents the robust lower bound of thermal conductivity for natural BLG. This value is temperature independent in the range of 250 K to 350 K which is the typical operating temperature range of most electronic devices.

This is indeed encouraging for the integration of graphene as a thermal-management material into modern integrated circuits, where exposure to polymeric materials such as photoresists during device fabrication makes surface contamination virtually inevitable[44, 45]. Although this realistic $\kappa$ falls below earlier theoretical expectations, its insensitivity to further surface scattering endows BLG with unique practical significance.

On the one hand, in micro/nanoscale thermal metrology, where conflicting results often arise from variations in sample suspension and surface cleanliness, BLG's stable $\kappa$ at the micrometer scale makes it a reference dataset for comparing thermal-transport measurements. On the other hand, for the future integration of graphene into CMOS-compatible thermal management architectures, our study provides the essential baseline experimental data. It highlights that engineers must account for this inevitable saturated value rather than relying on idealized theoretical values when designing graphene-based chip cooling solutions.

## Conclusions

In summary, we developed a sample preparation strategy that overcomes the bottleneck in fabricating 2D-material samples for suspended thermal bridge measurements. Utilizing this approach, we systematically measured the thermal properties of an extensive array of BLG samples under realistic surface contamination conditions. The BLG exhibits a length independent thermal conductivity of ~570 W/(m·K) over the length range of 1–30 μm in the temperature range of 250–350 K. MD simulations reveal that sub-monolayer surface coverage (~0.01 ML) is sufficient to saturate ZA phonon scattering, reaching a highly robust lower bound. These findings significantly deepen the fundamental understanding of phonon transport in graphene and provide a vital reference for its practical engineering applications.

## Methods

### Fabrication of the Suspended Micro-Bridge Devices.

The micro-bridge devices were fabricated utilizing standard microfabrication techniques. Initially, a 150-nm-thick $SiO_2$ layer and a 300-nm-thick $SiN_x$ layer were sequentially deposited onto a 500-μm-thick single-crystalline silicon wafer via low-pressure chemical vapor deposition (LPCVD). In the first photolithography step, a 5 nm / 85 nm Ti/Pt metal bilayer was deposited using electron-beam evaporation, followed by a standard lift-off process to define the electrode patterns. Subsequently, a second photolithography step, coupled with inductively coupled plasma (ICP) etching, was employed to pattern the top $SiO_2/SiN_x$ layers, thereby defining the geometries of the suspension beams and the silicon nitride islands. Finally, a third photolithography step was conducted, utilizing a combination of ICP and deep reactive-ion etching (DRIE) to selectively remove the underlying $SiO_2/SiN_x$ materials beneath the designated regions, achieving the complete release and suspension of the micro-bridge beams and the central $SiN_x$ islands.

**Preparation, Patterning, and Transfer of Graphene.**

High-quality natural graphite crystals were exfoliated via the mechanical cleavage method using adhesive tape and subsequently deposited onto a Si substrate coated with a 285-nm-thick $SiO_2$ layer. Prior to exfoliation, the $SiO_2$ surface was treated with oxygen plasma via RIE to enhance adhesion and render the surface hydrophilic. Optical microscopy was employed for initial screening, followed by micro-Raman spectroscopy to precisely determine the number of layers and evaluate the intrinsic lattice defects of the samples. For the patterning process, a 300-nm-thick PMMA layer, patterned via EBL, served as the etch mask. Subsequently, oxygen plasma RIE was utilized to etch the graphene, removing the unprotected regions to achieve well-defined geometrical patterns. To overcome the inherent challenges associated with transferring two-dimensional materials onto suspended micro-bridge devices, a polymer-assisted semi-dry nondestructive transfer technique was adopted. A temporary polymer protective layer of approximately 300 nm was spin-coated onto the patterned BLG. By default, PMMA 950K was utilized, whereas low-molecular-weight PMMA 50K or PPC were employed in control experiments. A 500-μm-thick elastic PDMS stamp was then attached to serve as the carrier film. Leveraging the physical intercalation of deionized water at the $SiO_2$ interface, the graphene, along with its polymer protective layer, was entirely delaminated from the Si substrate. Using a 2D material transfer platform, the BLG was precisely aligned and released across the suspended measurement platforms of the micro-bridge device. The samples were subsequently immersed in an acetone bath at 55 °C for 60 minutes, followed by multiple rinsing and substitution cycles with IPA to dissolve and eliminate the polymer layer and other residual impurities. Finally, to prevent capillary forces during the drying phase from collapsing or damaging the suspended graphene films and the ultrathin silicon nitride micro-bridges, the devices were dried immediately after cleaning using a carbon dioxide critical point dryer.

**Thermal Measurement Methodology.**

The in-plane thermal conductivity of BLG was characterized within a variable-temperature ultra-high vacuum (<$10^{-5}$ mbar) cryostat. This environment strictly eliminates parasitic heat losses arising from ambient thermal convection and residual gas conduction. During the measurements, a direct current of 15 μA was applied to the Heater to induce a stable, localized temperature rise. Simultaneously, a small alternating current of 0.5 μA at approximately 2000 Hz—with a slight frequency offset between the Heater and the Sensor—was applied to both platforms. Lock-in amplifiers were utilized to precisely monitor the electrical resistance of the coils. By calibrating the resistance as a function of the ambient temperature, the TCR for each coil was determined. This allowed for the accurate extraction of the Joule-heating-induced temperature rises on the Heater ($\Delta T_h$) and the Sensor ($\Delta T_S$). The apparent total thermal resistance of the sample ($R_{total}$) was subsequently calculated using the relation $R_{total} = (\Delta T_h^2 - \Delta T_s^2) / (Q\Delta T_s)$, where $Q$ represents the applied heating power.

**Sample Cleaning and Morphological Characterization.**

To systematically investigate the regulation mechanisms of the interfacial contamination layers, the transferred BLG samples were subjected to the following cleaning treatments: (1) $H_2$/Ar thermal annealing: Samples were annealed in a reducing mixed gas atmosphere at 400 °C for 4 hours. (2) $CO_2$ thermal annealing: Samples were annealed in a mildly oxidizing $CO_2$ atmosphere at 500 °C for 2 hours. (3) Vacuum laser cleaning: The suspended BLG surfaces were irradiated and cleaned using a 532 nm laser at a power of 3 mW for 300 s under high-vacuum conditions. (4) Ambient laser cleaning: Laser irradiation was performed using a 532 nm laser at a higher power of 12 mW for 240 s in ambient air. The surface topography and microscopic morphology of the samples, both prior to and following the cleaning protocols, were characterized utilizing SEM. To accurately assess the surface morphology, suspended BLG samples were fabricated on TEM-compatible silicon nitride holey membranes using the same preparation protocol. TEM characterization was then performed on the samples before and after the cleaning process. Furthermore, TEM coupled with EELS was employed to evaluate the atomic-scale structural evolution of the BLG surfaces induced by the cleaning processes. Specifically, we closely monitored the evolution of the energy-loss near-edge structure at the Carbon K-edge (~284 eV) and the desorption status of the Oxygen K-edge (~532 eV) characteristic peaks, serving as direct indicators of polymer residue elimination before and after the thermochemical treatments.

**Computational details.**

All molecular dynamics (MD) simulations were performed using the GPUMD package[46] with a neuroevolution potential (NEP) for graphene[47]. The NEP model was trained on the extensive carbon dataset established by Rowe et al.[48], which calculated via optB88-vdW DFT, yielding root-mean-square errors for graphene-type structures of 14.6 meV/atom for energy and 101.5 meV/Å for forces. Simulations utilized Bernal-stacked bilayer graphene (BLG, effective thickness 0.68 nm), with length and heat flux aligned along the zigzag y-direction. Periodic boundary conditions with x and y directions and a 1.0 fs time step were applied. Bulk thermal conductivity and finite-length ballistic transport were calculated via the homogeneous nonequilibrium molecular dynamics (HNEMD, 26.6 × 26.9 $nm^2$, Figure 4) and standard non-equilibrium molecular dynamics (NEMD, 26.6 × 53.9 $nm^2$), respectively. Each reported value is averaged over four independent MD simulations. To study surface condition effects, BLG models with 0% to 7.0% surface coverages (ratio of surface to single-layer carbon atoms) were constructed and relaxed under the NPT ensemble for 200 ps.

**HNEMD:** For the HNEMD-based calculations, the systems were first equilibrated in the canonical ensemble (constant number of particles *N*, volume *V,* and temperature *T*, NVT) at 300 K for 100 ps, followed by an 8 ns microcanonical ensemble (constant number of particles *N*, volume *V*, and total energy *E*, NVE) production run. A Langevin thermostat applied an external driving force ($F_e = 0.1$ μm$^{-1}$) to maintain a stable non-equilibrium state. Spectral thermal conductivity $k(\omega)$ was calculated via the Fourier transform of the accumulated virial-velocity correlation function[49]:

$$k(\omega) = \frac{2}{VTF_e}\int_0^{\infty} e^{i\omega t}K(t)dt\,, \tag{1}$$

where $V$ is the volume, $T$ represents the temperature, and $K(t) = \langle \sum \overrightarrow{W}_{i,0} \cdot \vec{v}_{i,t} \rangle$ is the accumulated virial-velocity correlation function. $\overrightarrow{W}_{i,0}$ is the virial tensor of atom *i* in time 0, $\vec{v}_{i,t}$ is the velocity tensor of atom *i* in time *t*.

**Corrected conductivity:** To compensate for the typical underestimation of lattice thermal conductivity by machine-learning potentials, we applied the correction framework by Wu et al.[50]:

$$\frac{1}{\kappa} = \frac{1}{\kappa_0} + \alpha\sigma_{\text{tot}}, \tag{2}$$

where $\kappa$ and $\kappa_0$ represent the error-affected and error-free thermal conductivities, respectively, and $\alpha$ is the slope. Since both the intrinsic NEP error and Langevin random forces are Gaussian distributions, the total force error scales accordingly according to Gaussian properties:

$$\sigma_{\text{tot}}^2 = \sigma_{\text{lan}}^2 + \sigma_{\text{NEP}}^2, \tag{3}$$

where $\sigma_{lan}$ is the additional random force from the Langevin thermostat, and $\sigma_{NEP}$ is 101.5 meV/Å. A correction factor of 1.55 was derived using the 0.01 ML model (Figure S6). This factor was uniformly multiplied to the calculated thermal conductivities across all coverages.

**NEMD:** For the NEMD-based calculations, 1 nm fixed regions were applied at both ends to eliminate rigid-body drift, adjacent to 10 nm heat source and sink regions. Following a 100 ps NVT relaxation, a ±10 K temperature difference was maintained between the source and sink for 1.0 ns in the NVE ensemble using Langevin thermostats. The ballistic thermal conductance $G(\omega)$ was calculated as follows[51]:

$$G(\omega) = \frac{2}{V\Delta T}\int_0^{\infty} e^{i\omega t}K(t)dt, \tag{4}$$

where *V* is the cross-sectional volume, and $\Delta T$ is the temperature difference between the source and the sink. The cross-sectional area was determined as the product of the system width and the effective thickness of the BLG.

**PDOS and MFP:** The phonon density of states $PDOS(\omega)$ is obtained from the Fourier transformation of the velocity auto-correlation function (VACF):

$$PDOS(\omega) = \int_{-\infty}^{\infty} e^{-i\omega t}\langle \vec{v}_{i,0} \cdot \vec{v}_{i,t}\rangle dt. \tag{5}$$

The spectral phonon mean free path $\lambda(\omega)$ was calculated from[49]:

$$\lambda(\omega) = \frac{\kappa(\omega)}{G(\omega)}. \tag{6}$$

## Supporting Information

Additional Raman characterization; temperature-dependent thermal resistance data; extraction of thermal conductivity for supported BLG; supplementary molecular dynamics results; measurement sensitivity and uncertainty analysis (PDF).

## Author Information

### Author Contributions

Y. G. and Y. D. jointly developed the patterning and transfer methods for two-dimensional materials, and performed sample preparation, thermal transport measurements, and part of the characterizations. X. Z., together with Y. G. and Y. D., designed and fabricated the thermal bridge devices. H. X. carried out the theoretical simulations. Z. T. and Y. G. prepared the TEM samples and performed the TEM characterizations. Z. Z. provided access to key experimental facilities, including SEM and EBL systems.

Y. G., Y. D., H. X., Z. T., F. D., W. L., G. Z., Y. X. and B. L. wrote the manuscript with input from all authors. G. Z., Y. X. and B. L. supervised the project and secured funding support. All authors discussed the results and commented on the manuscript.

**Notes**

The authors declare no competing financial interest.

## Acknowledgments

This project has been supported in part by Guangdong Provincial Quantum Science Strategic Initiative (Grant No. GDZX2502002), Guangdong Basic and Applied Basic Research Foundation (Grant No. 2023ZT10X010), Shenzhen Science and Technology Innovation Committee (Grant No. JCYJ20241202125413018, SYSRD20250529114001002), National Natural Science Foundation of China (Grant No. 12303096), Shenzhen Science and Technology Innovation Bureau (Grant No. 2025562891). The computational resources were supported by Centre for Computational Science and Engineering at Southern University of Science and Technology (SUSTech). We thank Dr. Xue Bai for insightful comments and suggestions that helped improve the quality of this work. We also thank Xin Peng, Dr. Hao Tian, and Dr. Linxin Zhai for their assistance in the development of the transfer method.

## References

1 Woon WY, Kasperovich A, Wen JR, Hu KK, Malakoutian M, Jhang JH*, et al.* Thermal management materials for 3D-stacked integrated circuits. *Nature Reviews Electrical Engineering* 2025, **2**(9)**:** 598-613.

2 van Erp R, Soleimanzadeh R, Nela L, Kampitsis G, Matioli E. Co-designing electronics with microfluidics for more sustainable cooling. *Nature* 2020, **585**(7824)**:** 211-216.

3 Wei B, Luo W, Du J, Ding Y, Guo Y, Zhu G*, et al.* Thermal interface materials: From fundamental research to applications. *SusMat* 2024, **4**(6)**:** e239.

4 Dou Z, Lei C, Wu K, Yu G. The development of thermal interface materials. *Nature Electronics* 2025, **8**(12)**:** 1146-1155.

5 Balandin AA, Ghosh S, Bao W, Calizo I, Teweldebrhan D, Miao F*, et al.* Superior Thermal Conductivity of Single-Layer Graphene. *Nano Letters* 2008, **8**(3)**:** 902-907.

6 Seol JH, Jo I, Moore AL, Lindsay L, Aitken ZH, Pettes MT*, et al.* Two-Dimensional Phonon Transport in Supported Graphene. *Science* 2010, **328**(5975)**:** 213-216.

7 Jang W, Chen Z, Bao W, Lau CN, Dames C. Thickness-Dependent Thermal Conductivity of Encased Graphene and Ultrathin Graphite. *Nano Letters* 2010, **10**(10)**:** 3909-3913.

8 Chen S, Moore AL, Cai W, Suk JW, An J, Mishra C*, et al.* Raman Measurements of Thermal Transport in Suspended Monolayer Graphene of Variable Sizes in Vacuum and Gaseous Environments. *ACS Nano* 2010, **5**(1)**:** 321-328.

9 Wang Z, Xie R, Bui CT, Liu D, Ni X, Li B*, et al.* Thermal Transport in Suspended and Supported Few-Layer Graphene. *Nano Letters* 2010, **11**(1)**:** 113-118.

10 Ghosh S, Bao W, Nika DL, Subrina S, Pokatilov EP, Lau CN*, et al.* Dimensional crossover of thermal transport in few-layer graphene. *Nature Materials* 2010, **9**(7)**:** 555-558.

11 Pettes MT, Jo I, Yao Z, Shi L. Influence of Polymeric Residue on the Thermal Conductivity of Suspended Bilayer Graphene. *Nano Letters* 2011, **11**(3)**:** 1195-1200.

12 Jang W, Bao W, Jing L, Lau CN, Dames C. Thermal conductivity of suspended few-layer

graphene by a modified T-bridge method. *Applied Physics Letters* 2013, **103**(13): 133102.

13 Sadeghi MM, Jo I, Shi L. Phonon-interface scattering in multilayer graphene on an amorphous support. *PNAS* 2013, **110**(41): 16321-16326.

14 Xu X, Pereira LFC, Wang Y, Wu J, Zhang K, Zhao X, *et al.* Length-dependent thermal conductivity in suspended single-layer graphene. *Nature Communications* 2014, **5**(1): 3689.

15 Wang H, Kurata K, Fukunaga T, Ago H, Takamatsu H, Zhang X, *et al.* Simultaneous measurement of electrical and thermal conductivities of suspended monolayer graphene. *Journal of Applied Physics* 2016, **119**(24): 244306.

16 Mercado E, Anaya J, Kuball M. Impact of Polymer Residue Level on the In-Plane Thermal Conductivity of Suspended Large-Area Graphene Sheets. *ACS Applied Materials & Interfaces* 2021, **13**(15): 17910-17919.

17 Liu C, Lu P, Chen W, Zhao Y, Chen Y. Phonon transport in graphene based materials. *Physical Chemistry Chemical Physics* 2021, **23**(46): 26030-26060.

18 Li J, Zheng J, Zhai L, Liu P, Qing F, Li X, *et al.* Layer-Dependent Thermal Conductivity of Graphene toward Thermal Management Applications. *ACS Applied Materials & Interfaces* 2026, **18**(7): 11764-11773.

19 Gu X, Wei Y, Yin X, Li B, Yang R. Colloquium: Phononic thermal properties of two-dimensional materials. *Reviews of Modern Physics* 2018, **90**(4): 041002.

20 Han Z, Ruan X. Thermal conductivity of monolayer graphene: Convergent and lower than diamond. *Physical Review B* 2023, **108**(12): L121412.

21 Wang J, Zhu L, Chen J, Li B, Thong JTL. Suppressing Thermal Conductivity of Suspended Tri-layer Graphene by Gold Deposition. *Advanced Materials* 2013, **25**(47): 6884-6888.

22 Vallabhaneni AK, Singh D, Bao H, Murthy J, Ruan X. Reliability of Raman measurements of thermal conductivity of single-layer graphene due to selective electron-phonon coupling: A first-principles study. *Physical Review B* 2016, **93**(12): 125432.

23 Shi L, Li D, Yu C, Jang W, Kim D, Yao Z, *et al.* Measuring Thermal and Thermoelectric Properties of One-Dimensional Nanostructures Using a Microfabricated Device. *Journal of Heat Transfer* 2003, **125**(5): 881-888.

24 Li D, Wu Y, Kim P, Shi L, Yang P, Majumdar A. Thermal conductivity of individual silicon nanowires. *Applied Physics Letters* 2003, **83**(14): 2934-2936.

25 Jo I, Pettes MT, Kim J, Watanabe K, Taniguchi T, Yao Z, *et al.* Thermal Conductivity and Phonon Transport in Suspended Few-Layer Hexagonal Boron Nitride. *Nano Letters* 2013, **13**(2): 550-554.

26 Guo J, Huang Y, Wu X, Wang Q, Zhou X, Xu X, *et al.* Thickness-Dependent In-Plane Thermal Conductivity and Enhanced Thermoelectric Performance in p-Type ZrTe5 Nanoribbons. *physica status solidi (RRL) – Rapid Research Letters* 2018, **13**(3): 1800529.

27 Xie Y, Karamati A, Wang X. Transient electro-thermal technique for measuring the thermal diffusivity/conductivity of 1D/2D materials: from mm down to atomic scale thickness. *Thermo-X* 2025, **1**: 202503.

28 Li H, Wu J, Huang X, Yin Z, Liu J, Zhang H. A Universal, Rapid Method for Clean Transfer

of Nanostructures onto Various Substrates. *ACS Nano* 2014, **8**(7): 6563-6570.

29 Malard LM, Pimenta MA, Dresselhaus G, Dresselhaus MS. Raman spectroscopy in graphene. *Physics Reports* 2009, **473**(5-6): 51-87.

30 Ferrari AC, Meyer JC, Scardaci V, Casiraghi C, Lazzeri M, Mauri F*, et al.* Raman Spectrum of Graphene and Graphene Layers. *Physical Review Letters* 2006, **97**(18): 187401.

31 Bae M-H, Li Z, Aksamija Z, Martin PN, Xiong F, Ong Z-Y*, et al.* Ballistic to diffusive crossover of heat flow in graphene ribbons. *Nature Communications* 2013, **4**(1): 1734.

32 Lindsay L, Broido DA, Mingo N. Flexural phonons and thermal transport in multilayer graphene and graphite. *Physical Review B* 2011, **83**(23): 235428.

33 Fugallo G, Cepellotti A, Paulatto L, Lazzeri M, Marzari N, Mauri F. Thermal Conductivity of Graphene and Graphite: Collective Excitations and Mean Free Paths. *Nano Letters* 2014, **14**(11): 6109-6114.

34 Lin Y-C, Lu C-C, Yeh C-H, Jin C, Suenaga K, Chiu P-W. Graphene Annealing: How Clean Can It Be? *Nano Letters* 2011, **12**(1): 414-419.

35 Gong C, Floresca HC, Hinojos D, McDonnell S, Qin X, Hao Y*, et al.* Rapid Selective Etching of PMMA Residues from Transferred Graphene by Carbon Dioxide. *The Journal of Physical Chemistry C* 2013, **117**(44): 23000-23008.

36 Zhang J, Jia K, Lin L, Zhao W, Quang HT, Sun L*, et al.* Large-Area Synthesis of Superclean Graphene via Selective Etching of Amorphous Carbon with Carbon Dioxide. *Angew Chem Int Ed Engl* 2019, **58**(41): 14446-14451.

37 Jia Y, Gong X, Peng P, Wang Z, Tian Z, Ren L*, et al.* Toward High Carrier Mobility and Low Contact Resistance: Laser Cleaning of PMMA Residues on Graphene Surfaces. *Nano-Micro Letters* 2016, **8**(4): 336-346.

38 Mondal A, Biswas C, Park S, Cha W, Kang S-H, Yoon M*, et al.* Low Ohmic contact resistance and high on/off ratio in transition metal dichalcogenides field-effect transistors via residue-free transfer. *Nature Nanotechnology* 2023, **19**(1): 34-43.

39 Schweizer P, Dolle C, Dasler D, Abellán G, Hauke F, Hirsch A*, et al.* Mechanical cleaning of graphene using in situ electron microscopy. *Nature Communications* 2020, **11**(1): 1743.

40 Chen J, Zhang G, Li B. Substrate coupling suppresses size dependence of thermal conductivity in supported graphene. *Nanoscale* 2013, **5**(2): 532-536.

41 Ye Z-Q, Cao B-Y, Guo Z-Y. Study on thermal characteristics of phonons in graphene. *Acta Physica Sinica* 2014, **63**(15): 154704.

42 Lindsay L, Broido DA, Mingo N. Flexural phonons and thermal transport in graphene. *Physical Review B* 2010, **82**(11): 115427.

43 Qiu B, Ruan X. Reduction of spectral phonon relaxation times from suspended to supported graphene. *Applied Physics Letters* 2012, **100**(19): 193101.

44 Li Z, Wang Y, Kozbial A, Shenoy G, Zhou F, McGinley R*, et al.* Effect of airborne contaminants on the wettability of supported graphene and graphite. *Nat Mater* 2013, **12**(10): 925-931.

45 Tang Z, Garcia R. Angstrom-Scale Water Layer Structure on van der Waals Materials Probed by 3D Atomic Force Microscopy: From Acidic to Alkaline Aqueous Solutions. *Langmuir* 2026, **42**(7): 5608-5616.

46 Xu K, Bu H, Pan S, Lindgren E, Wu Y, Wang Y, *et al.* GPUMD 4.0: A high-performance molecular dynamics package for versatile materials simulations with machine-learned potentials. *Materials Genome Engineering Advances* 2025, **3:** e70028.

47 Fan Z, Xiao Y, Wang Y, Ying P, Chen S, Dong H. Combining linear-scaling quantum transport and machine-learning molecular dynamics to study thermal and electronic transports in complex materials. *J Phys: Condens Matter* 2024, **36**(24)**:** 245901.

48 Rowe P, Deringer VL, Gasparotto P, Csányi G, Michaelides A. An accurate and transferable machine learning potential for carbon. *The Journal of Chemical Physics* 2020, **153**(3)**:** 034702.

49 Fan Z, Dong H, Harju A, Ala-Nissila T. Homogeneous nonequilibrium molecular dynamics method for heat transport and spectral decomposition with many-body potentials. *Physical Review B* 2019, **99**(6)**:** 064308.

50 Wu X, Zhou W, Dong H, Ying P, Wang Y, Song B, *et al.* Correcting force error-induced underestimation of lattice thermal conductivity in machine learning molecular dynamics. *The Journal of Chemical Physics* 2024, **161**(1)**:** 014103.

51 Fan Z, Pereira LFC, Hirvonen P, Ervasti MM, Elder KR, Donadio D, *et al.* Thermal conductivity decomposition in two-dimensional materials: Application to graphene. *Physical Review B* 2017, **95**(14)**:** 144309.

# Supporting Information

# *Lower Bound of Thermal Conductivity in Graphene*

## 1. Material characterization details

### 1.1 Supplementary Raman spectra of the samples

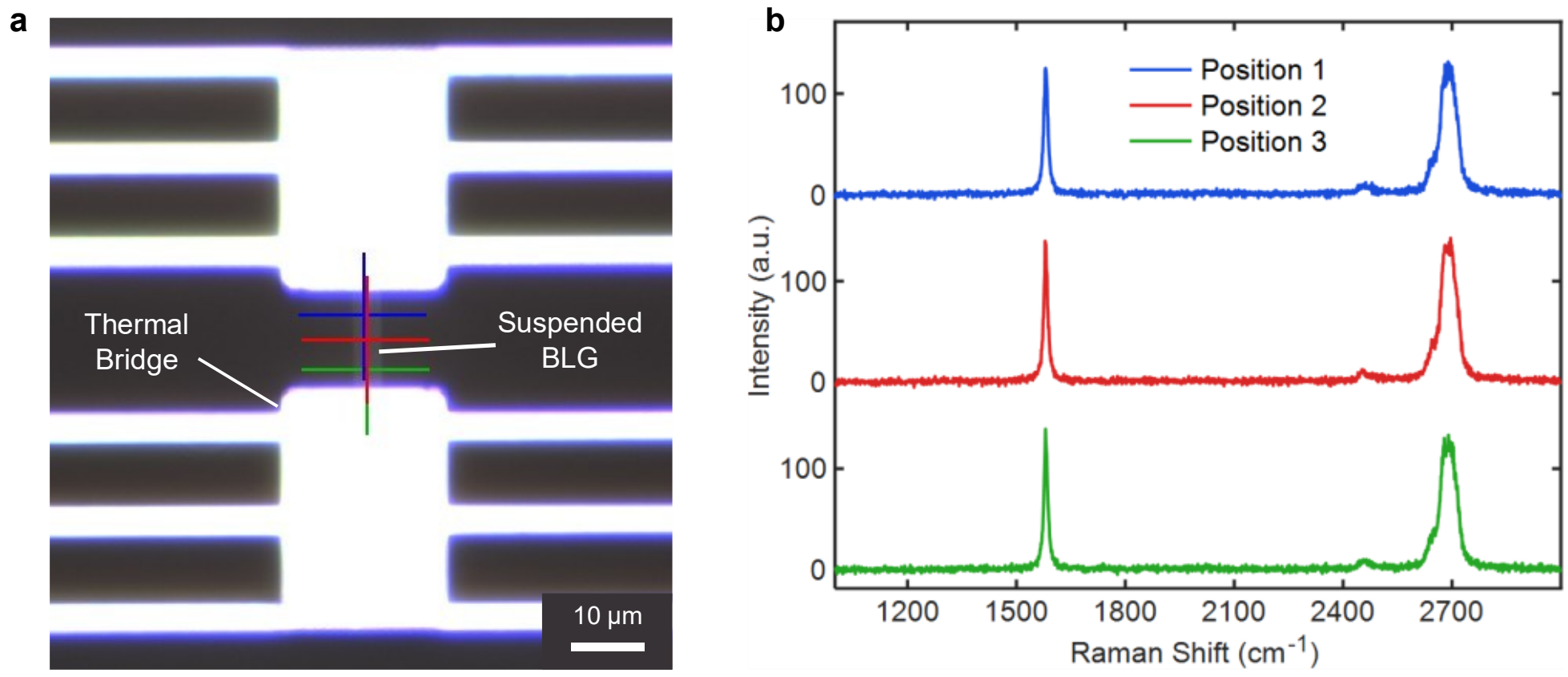

**Figure S1. Typical optical micrograph and Raman spectra of a suspended BLG sample. a**, Optical micrograph of a suspended BLG sample (15 μm in length and 4 μm in width). An exceptionally high illumination intensity was employed due to the intrinsically low optical contrast of BLG. **b**, Raman spectra of the BLG sample shown in **a**. The spectra in different colors were acquired from the positions indicated by the correspondingly colored crosses in **a**. The almost imperceptible D band (~1350 $cm^{-1}$) indicates an extremely low defect density in the sample.

## 1.2 Detailed data of total thermal resistance versus temperature for the BLG

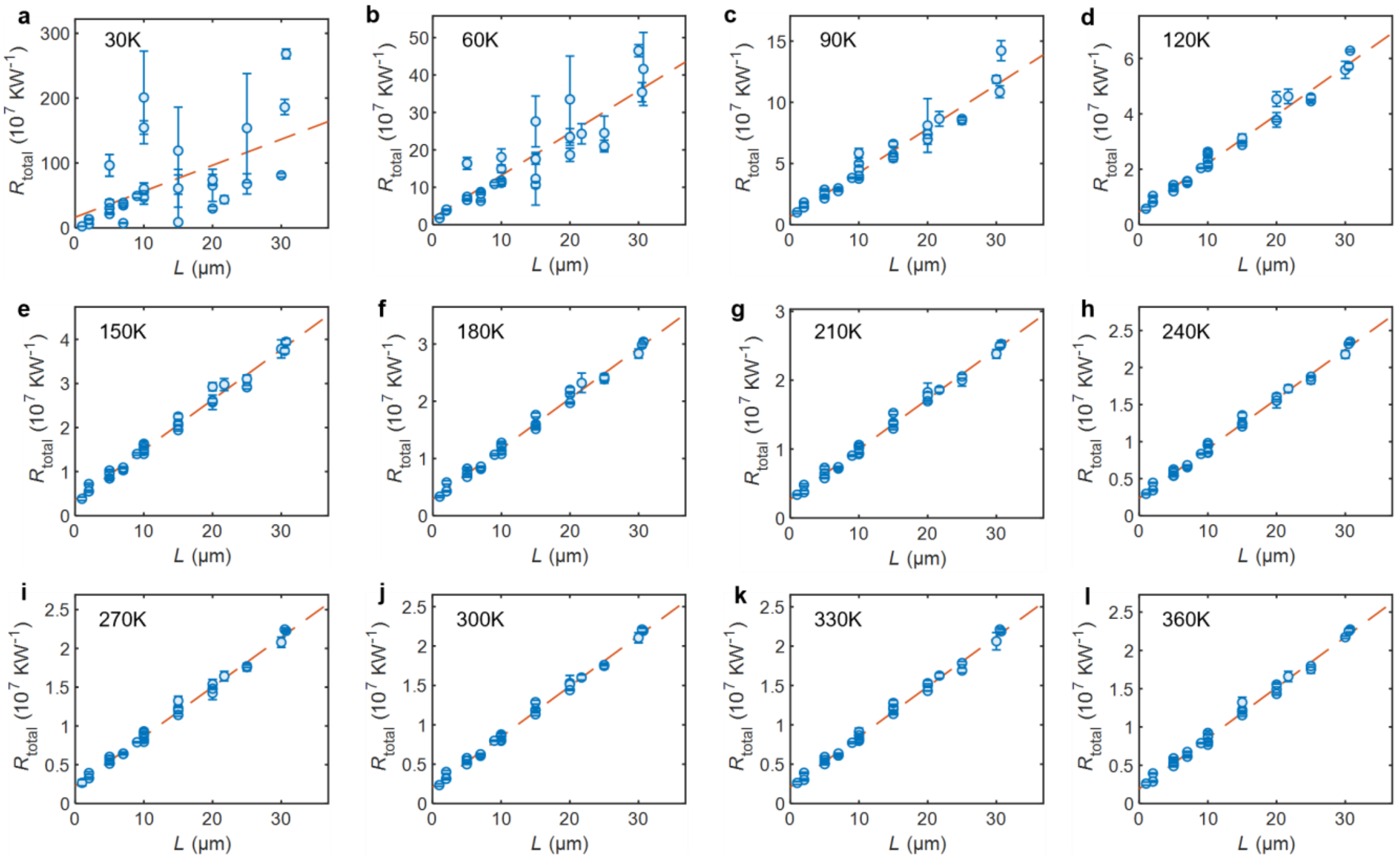

**Figure S2. Relationship between the total thermal resistance ($R_{total}$) and length ($L$) of BLG at various temperatures.** **a-l**, $R_{total}$ versus $L$ for BLG samples at temperatures ranging from 30 K to 360 K. All BLG samples have a uniform width ($W$) of 4 μm and were prepared via a PMMA-assisted transfer method. Due to the larger measurement error of thermal resistance at 30 K, the linear fitting quality is relatively poor. However, at 90 K and above, the coefficients of determination are all greater than 0.95.

## 1.3 Raman characterization of the samples before and after cleaning

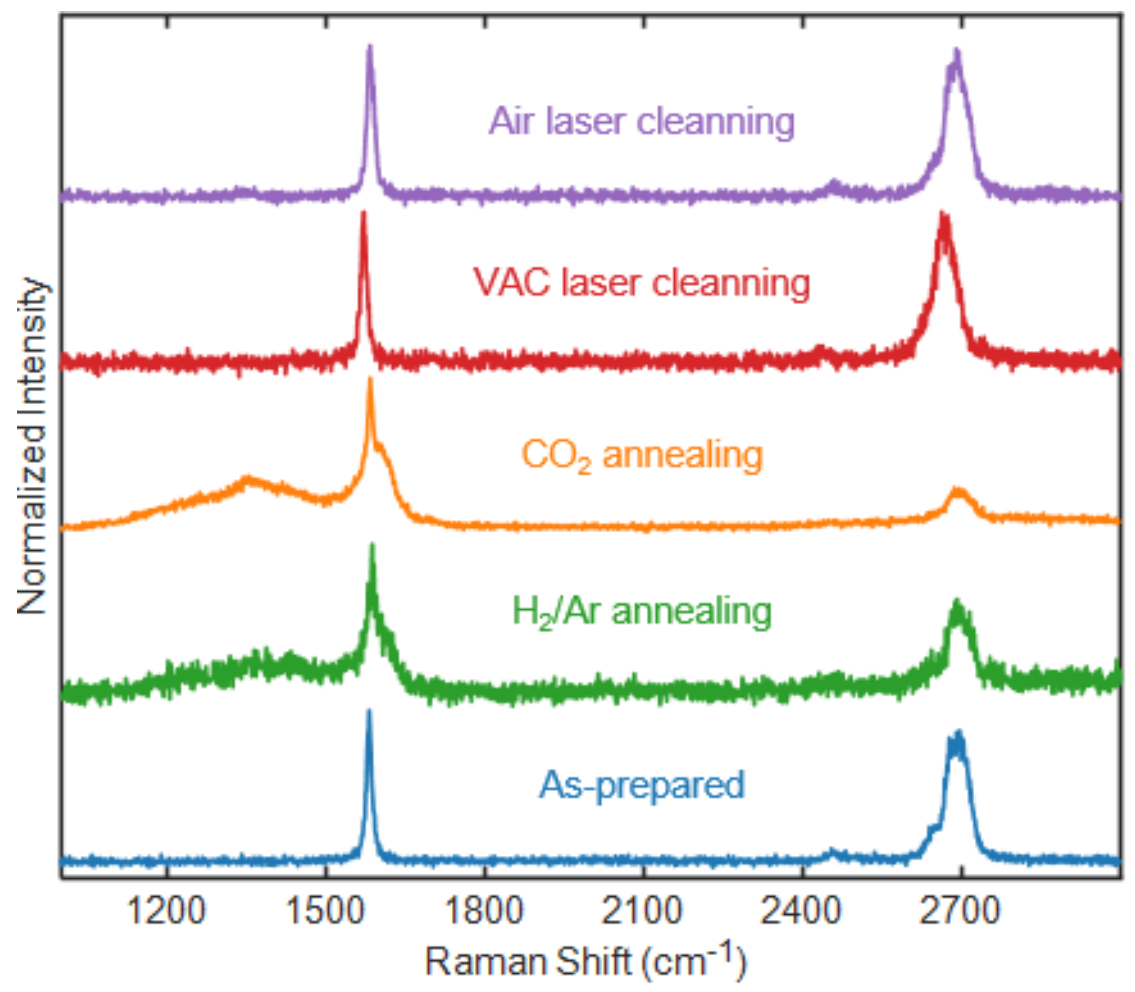

**Figure S3. Comparison of Raman spectra of the samples before and after different cleaning treatments.**

As shown in Figure S3, the D band of the graphene sample before cleaning is almost imperceptible because high-quality natural graphite was mechanically exfoliated to prepare the graphene. After annealing in $H_2$/Ar at 400 °C for 4 h, a very broad D band emerges. However, we do not attribute this feature to the oxidation of graphene. Instead, it is likely a Raman signal arising from the dehydrogenation and deoxygenation of surface PMMA residues during the thermal annealing process, which forms carbon chains of varying lengths[1-4]. A similar mechanism is expected to be responsible for the very broad D band observed after $CO_2$ annealing at 500 °C for 2 h. On the other hand, the sample cleaned by laser in a vacuum shows no obvious D band. This, however, does not necessarily indicate that the sample surface is absolutely clean. the absence of the broad D band (in contrast to the thermally annealed samples) is likely due to the distinct cleaning mechanism of laser cleaning[5]. Laser cleaning performed under ambient conditions led to a weak D band, which could be attributed to slight oxidation of the sample caused by either the relatively high laser power (12 mW) or the prolonged exposure time (240 s).

### 1.4 Thermal conductivity calculation of supported BLG

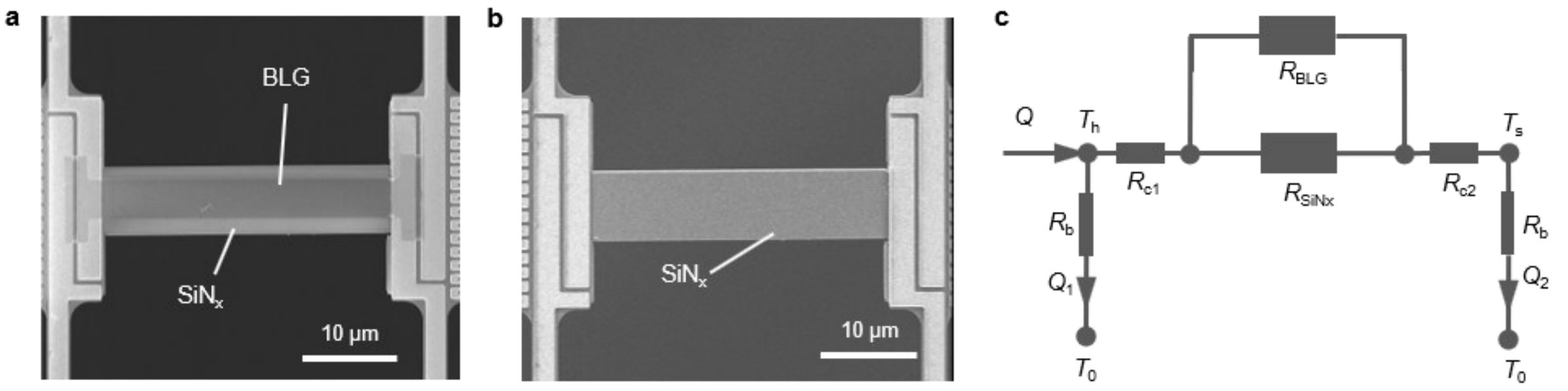

**Figure S4. a**, SEM image of the graphene sample on the silicon nitride ($SiN_x$) substrate. **b**, SEM image of the bare $SiN_x$ substrate. **c**, Equivalent thermal circuit of the device.

In this experiment, thermal measurements were conducted on both the silicon nitride ($SiN_x$) substrate with the transferred BLG sample (as shown in Figure S4a) and the bare $SiN_x$ substrate (as shown in Figure S4b) using a suspended microbridge method. The heat transport process of the system can be described by a one-dimensional equivalent thermal circuit model (see Figure S4c). Under steady-state heating conditions, the heat $Q$ generated by the heating electrode is injected into the heating end (at temperature $T_\mathrm{h}$), and a portion of the heat, $Q_1$, dissipates to the ambient heat sink (at temperature $T_0$) through the supporting beam (with thermal resistance $R_\mathrm{b}$). Another portion of the heat flows through the central sample bridge region to the sensor end(at temperature $T_\mathrm{s}$), and subsequently dissipates as heat $Q_2$ through the supporting beam on the opposite end.

First, the total thermal resistance $R_\mathrm{total}$ of the bridge region is measured using the method described in the main text. Here, $R_\mathrm{total}$ consists of the contact thermal resistances ($R_\mathrm{c1}$ and $R_\mathrm{c2}$) and the parallel thermal resistance of the sample region. The sample region comprises the top BLG sample (with thermal resistance $R_\mathrm{BLG}$) and the underlying $SiN_x$ substrate (with thermal resistance $R_\mathrm{SiNx}$). To simplify the calculation, we combine the contact thermal resistances on both sides into a total contact thermal resistance $R_\mathrm{c}$. Therefore, the total thermal resistance of the bridge region covered with the BLG sample can be expressed as:

$$R_{\text{total}} = R_{\text{c}} + \left(\frac{1}{R_{\text{SiNx}}} + \frac{1}{R_{\text{BLG}}}\right)^{-1}. \tag{S1}$$

To accurately extract the intrinsic thermal resistance $R_{\text{BLG}}$ of the BLG, it is necessary to decouple the parasitic heat conduction of the substrate and the contact thermal resistance. First, the total contact thermal resistance $R_{\text{c}}$ can be obtained through fitting using the intercept method; specifically, by measuring and fitting the total thermal resistance of samples with different bridge lengths, the extrapolated intercept value at a length of zero yields $R_{\text{c}}$. Secondly, to eliminate the influence of the background thermal resistance of the $SiN_x$ substrate, after completing the aforementioned measurements, we utilized reactive ion etching (RIE) technology to completely remove the BLG sample on the bridge region. The total thermal resistance of the bare $SiN_x$ bridge, $R_{\text{bare}}$, was then measured again under identical environmental conditions. At this point, since the parallel channel $R_{\text{BLG}}$ is eliminated, the relationship simplifies to:

$$R_{\text{bare}} = R_{\text{c}} + R_{\text{SiN}_x}. \tag{S2}$$

From the above equation, the intrinsic thermal resistance of the substrate is $R_{\text{SiN}_x} = R_{\text{bare}} - R_{\text{c}}$. By combining the above equations and removing the contributions of the contact thermal resistance and the parallel thermal resistance of the substrate, the thermal resistance of the BLG sample, $R_{\text{BLG}}$, can be analytically expressed using the experimentally measured $R_{\text{total}}$, $R_{\text{bare}}$, and $R_{\text{c}}$:

$$\frac{1}{R_{\text{BLG}}} = \frac{1}{R_{\text{total}} - R_{\text{c}}} - \frac{1}{R_{\text{bare}} - R_{\text{c}}}. \tag{S3}$$

Further rearrangement yields:

$$R_{\text{BLG}} = \frac{(R_{\text{total}} - R_{\text{c}})(R_{\text{bare}} - R_{\text{c}})}{R_{\text{bare}} - R_{\text{total}}}. \tag{S4}$$

Finally, according to Fourier's law of heat conduction, the in-plane thermal conductivity $\kappa$ of the BLG can be calculated from the geometric dimensions of the sample:

$$\kappa = \frac{L}{R_{\text{BLG}} W h}, \tag{S5}$$

where $L$, $W$, and $h$ represent the effective suspended length, width, and thickness of the BLG sample, respectively.

## 2. Molecular dynamics simulation details

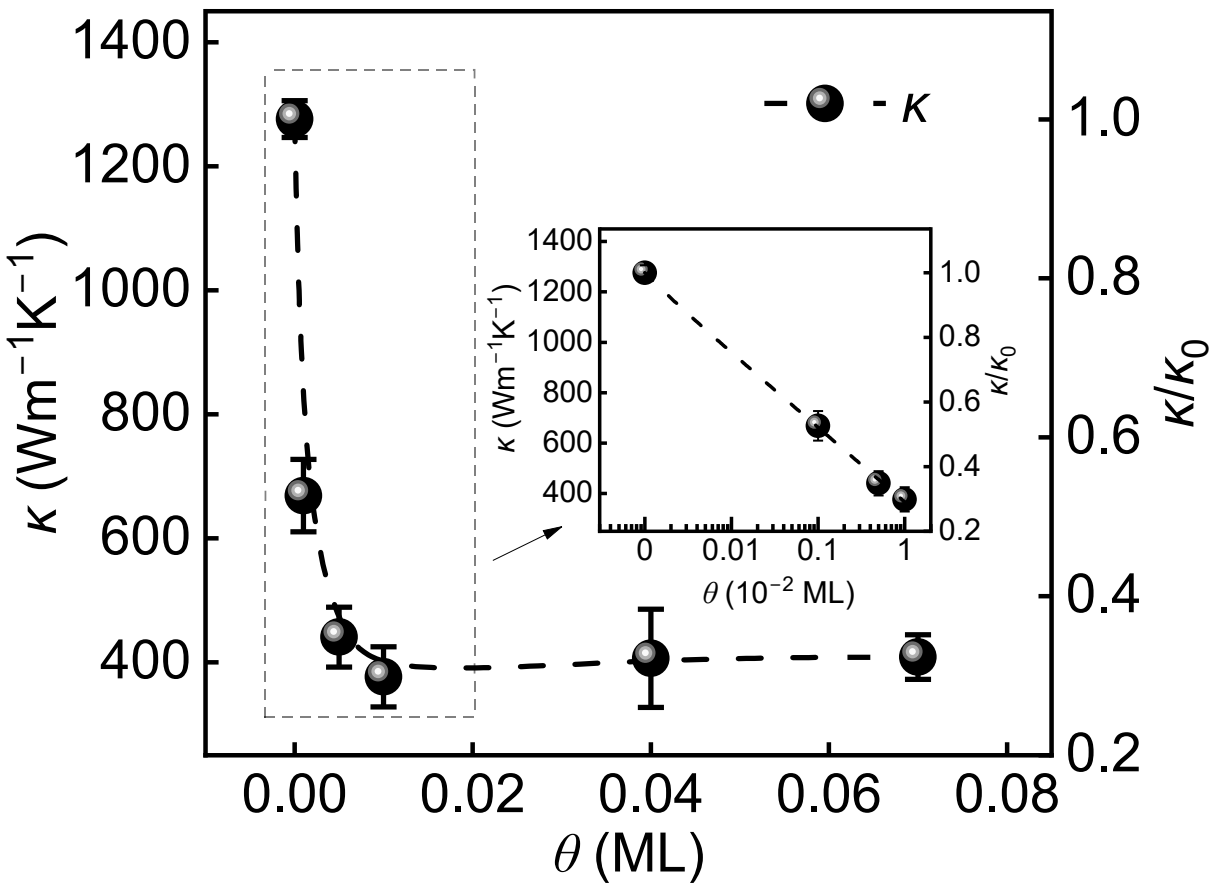

**Figure S5. Thermal conductivity at different coverages, the inset shows the fitted curve for low coverages.**

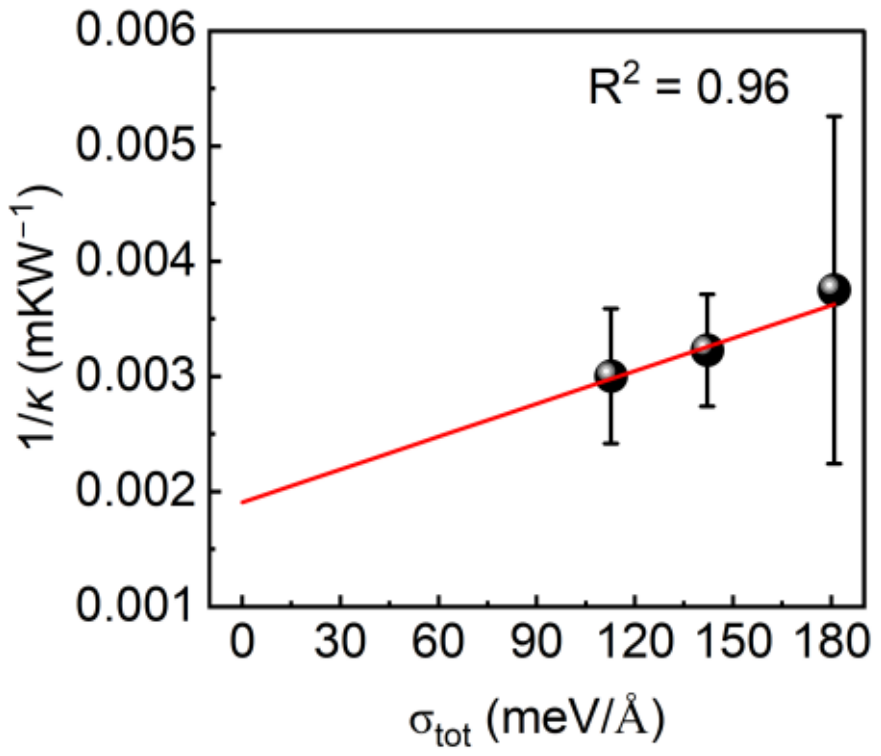

**Figure S6. The MD calculated the inverse thermal conductivity as a function of the total force error $\sigma_{\mathrm{tot}}$.** The red line represents a linear fit with the intercept at $\sigma_{\mathrm{tot}}=0$ corresponding to the corrected thermal conductivity value.

## 3. Uncertainty analysis

### 3.1 Thermal conductance measurement sensitivity characterization

For the thermal conductivity measurement, the sensitivity of thermal conductance measurement is determined by the minimum or noise-equivalent sample thermal conductance that can be measured using the microdevice. In our measurements, the temperatures of the heating/sensing membranes are measured by the electrical resistance of the platinum thermometers. from Eq S6, the noise-equivalent thermal conductance ($NEG_s$) of the sample is proportional to the noise-equivalent temperature rise (NET) of the sensing membrane, and NET is further related to the noise equivalent resistance (NER),

$$NEG_{\mathrm{s}} = G_{\mathrm{b}}\frac{NET}{\Delta T_{\mathrm{h}} - \Delta T_{\mathrm{s}}}, \tag{S6}$$

$$NET = \frac{NER/R_{\mathrm{s}}}{TCR}, \tag{S7}$$

where $G_{\mathrm{b}}$ is the thermal conductance of supporting beams, and TCR is the temperature coefficient of resistance of the integrated metal thermometers. For the resistance measurement method using a lock-in amplifier,

$$\frac{NER}{R_{\mathrm{s}}} = \frac{\delta_V}{V} + \frac{\delta_i}{i}, \tag{S8}$$

where $\delta_v$ and $\delta_i$ are the noises in the ac voltage measurement and that of the current source, while $V$ and $i$ represent the ac voltage drop across the sensor and the ac excitation current, respectively. At 300 K, $\delta_V$ is dominated by the Johnson noise， $\delta_V = \sqrt{4k_BTR_s\Delta f} \sim 4.3nV$(at 300 K a noise bandwidth of 0.25 Hz, $R_s$~5000 Ω),

$$\frac{\delta_V}{V} = \frac{4.3\times10^{-9}\ \mathrm{V}}{2.5\times10^{-3}\ \mathrm{V}} = 1.72\times10^{-6}. \tag{S9}$$

The current is provided by Keithley 6221 current source, to ensure the temperature rise of the heat membrane is ~ 5 K, set the DC current at 15 μA. according the User's Manual of 6221,

$$\frac{\delta_i}{i} = \frac{0.5\times10^{-9}\ \mathrm{A}}{1.5\times10^{-5}\ \mathrm{A}} = 3.33\times10^{-5}. \tag{S10}$$

Therefore,

$$\frac{NER}{R_{\rm s}} \approx 3.51 \times 10^{-5}. \tag{S11}$$

The TCR of the Pt resistance thermometer (PRT) is ~7×$10^{-3}$ $K^{-1}$ and ~2×$10^{-3}$ $K^{-1}$ about at 30 K and 300 K respectively. Thus the NET of the lock-in measurement is ~5.0 mK and ~17.5 mK at 30 K and 300 K, respectively, which is below the nominal temperature fluctuation of the Janis cryostat. On the other hand, the $G_{\rm b}$ has been calculated from the measured temperature rise of the PRT. In our measurement, $G_{\rm b}$ is ~0.5×$10^{-7}$ W/K and ~1.6×$10^{-7}$ W/K at 30 K and 300 K, so the noise-equivalent thermal conductance is about 5.0×$10^{-11}$ W/K and 5.6×$10^{-10}$ W/K at 30 K and 300 K respectively. The minimum possible thermal conductance of graphene would be 6.7×$10^{-10}$ W/K and 4.5×$10^{-8}$ W/K at 30 K and 300 K (for a BLG sample 30 μm in length and 4 μm in width) respectively. This is in the range of the measurement sensitivity.

### 3.2 Influence of thermal radiation

In the thermal conductivity measurement, we focus on the heat conducted by graphene. One important error source of the measurement may be the heat transfer between the two membranes via radiation and air conduction. Here, we conduct a detailed assessment for this question.

The radiation thermal conductance can be estimated as

$$G_{\rm h-s_radiation} = \frac{\varepsilon_{\rm eff}\sigma\left(T_{\rm h}^4 - T_{\rm s}^4\right)F_{\rm h-s}A_m}{T_{\rm h} - T_{\rm s}}, \tag{S12}$$

where $\sigma$ is the Stefan–Boltzmann constant, $\varepsilon_{\rm eff}$ is the effective emissivity between the membranes, and $F_{\rm h\text{-}s}$ and $A_m$ are the view factor between the two adjacent membranes and the surface area of one membrane, respectively. According to device design, the $F_{h-s}A = 9.63 \times 10^{-12}\ {\rm m}^2$, and the $G_{\rm h-s_radiation} = 7.54 \times 10^{-14}\ {\rm W/K}$ and $1.04 \times 10^{-10}\ {\rm W/K}$ at 30 K and 360 K respectively.

On the other hand, because of the radiation heat loss from the sample surface to the surroundings, the heat flux received by the sensing side membrane is always smaller than that transmitted from the heating side, and for the same aspect ratio, the radiation heat loss will increase with increasing sample length. As such, we need to carefully estimate the radiation heat loss for graphene samples with long suspended length.

The radiation thermal conductance of graphene can be estimated by Stefan–Boltzmann law,

$$G_{\rm gra_radiation} = \frac{2\varepsilon_{\rm gra}\sigma\left(T_{\rm gra}^4 - T_{\rm s}^4\right)F_{\rm gra-s}A_{\rm gra}}{T_{\rm gra} - T_{\rm s}}, \tag{S13}$$

where $\varepsilon_{\rm gra}$ is the emissivity of graphene, $F_{\rm gra-s}$ are the view factor between the graphene and the radiation shield, and $A_{\rm gra}$ is the surface area of graphene. 2 indicate heat from the graphene is exchanged via thermal radiation from its top and bottom surfaces. For the longest graphene sample measured in this work (length: 30 μm，width: 4 μm), the radiation thermal conductance caused by radiation of graphene is 1.88×$10^{-12}$ W/K and 2.59×$10^{-9}$ W/K at 30 K and 360 K respectively. Strictly speaking, under the assumptions of uniform graphene temperature, perfect emission ($\varepsilon_{\rm gra}$ = 1), and complete view factor ($F_{\rm gra-s}$ = 1), the conductive error due to outward thermal radiation from the graphene is lower than the computed value above, thus the radiation heat loss is not important for the long samples examined in this paper.

### 3.3 Influence of thermal convection

The thermal conductance by air can be recorded as $G_{\text{h-s_air}}$, which is calculated by

$$G_{\text{h-s_air}} = \kappa_{\text{air}} \frac{2A_m}{L}, \tag{S14}$$

where $\kappa_{air}$ is the thermal conductivity of the residual air molecules in the evacuated cryostat, $L$ are the distance between the two membranes. The measurement system we used can achieve a vacuum level of ~$10^{-5}$ mbar at room temperature, while the mean free path of air molecules is of the order of ~10 m and is much larger than D (Maximum value is 30 μm). According to the kinetic theory,

$$\kappa_{\text{air}} = \frac{CvL}{3}\text{, and } G_{\text{h-s_air}} = \frac{2CvA_m}{3}, \tag{S15}$$

where $C$ and $v$ are the heat capacity and velocity of air molecules, it can be estimated that $G_{\text{h-s_air}} \sim 10^{-10}\ W/K$ at room temperature. We also measured the thermal conductance due to air conduction and radiation using a bare device without graphene bridging the two membranes. No signal above the noise level can be detected by the sensing membrane as the temperature of the heating membrane is raised. The measurement confirms that air conduction and radiation do not introduce noticeable errors in our measurements.

### 3.4 Sample preparation error

Due to the geometric errors inevitably introduced during device fabrication, sample patterning, and transfer processes, the total measurement uncertainty of the in-plane thermal conductivity $\kappa$ of BLG in this study is rigorously evaluated using standard error propagation theory. Because the intrinsic thermal resistance is extracted via the intercept method, the statistical fitting of a large number of independent samples provides a robust foundation for assessing measurement errors.

Based on the earlier derivation, the thermal conductivity is calculated from the slope $m = \frac{dR_{\text{BLG}}}{dL}$ of the linear fit between thermal resistance and suspended length:

$$\kappa = \frac{1}{mWh}. \tag{S16}$$

The relative standard error of the thermal conductivity, $\frac{\Delta\kappa}{\kappa}$, is synthesized from three orthogonal components:

$$\frac{\Delta\kappa}{\kappa} = \sqrt{\left(\frac{\Delta m}{m}\right)^2 + \left(\frac{\Delta W}{W}\right)^2 + \left(\frac{\Delta h}{h}\right)^2}. \tag{S17}$$

**Statistical Uncertainty of the Fitted Slope ($\Delta m$)**: This constitutes the dominant source of uncertainty in our measurements. When extracting the slope $m$, the linear regression of $R_{\text{BLG}}$ versus $L$ not only decouples the contact thermal resistance but also yields a standard error of the slope that naturally captures the random sample-to-sample variations introduced during micro/nanofabrication and transfer processes. This encompasses localized random deviations in the lithographically defined suspended length $L$ and width $W$, as well as unavoidable subtle fluctuations among different devices. Consequently, these random geometric errors are accounted for within the fitting error, precluding the need for redundant calculations.

**Systematic Error in Average Width Measurement ($\Delta W$)**: Although random width fluctuations are included in $\Delta m$, SEM image measurements reveal that the sample widths are distributed between 4.0 and 4.1 μm. The width variation introduces a calculated relative error of 0.0122.

**Uncertainty in Thickness Definition ($\Delta h$)**: We reliably confirm the bilayer nature of the samples by performing Lorentzian fitting on the 2D Raman band, and accordingly, we adopt the standard BLG thickness of 0.68 nm. Since the intrinsic thermal conductivity of surface contamination is significantly lower than that of graphene, we do not consider the apparent thickness increase caused by such contamination as contributing to the effective heat conduction cross-section. Consequently, the relative error introduced by the thickness term is considered to be 0. Ultimately, through the combined propagation of the aforementioned systematic and statistical errors, the thermal conductivity of the bilayer graphene at 300 K is determined to be 572.7 ± 10.3 W/(m·K).

## References

1. Lin Y-C, Lu C-C, Yeh C-H, Jin C, Suenaga K, Chiu P-W. Graphene Annealing: How Clean Can It Be? Nano Letters 2011, 12(1): 414-419.

2. Tripathi M, Mittelberger A, Mustonen K, Mangler C, Kotakoski J, Meyer JC, et al. Cleaning graphene: Comparing heat treatments in air and in vacuum. physica status solidi (RRL) – Rapid Research Letters 2017, 11(8): 1700124.

3. Gong C, Floresca HC, Hinojos D, McDonnell S, Qin X, Hao Y, et al. Rapid Selective Etching of PMMA Residues from Transferred Graphene by Carbon Dioxide. The Journal of Physical Chemistry C 2013, 117(44): 23000-23008.

4. Zhang J, Jia K, Lin L, Zhao W, Quang HT, Sun L, et al. Large-Area Synthesis of Superclean Graphene via Selective Etching of Amorphous Carbon with Carbon Dioxide. Angew Chem Int Ed Engl 2019, 58(41): 14446-14451.

5. Jia Y, Gong X, Peng P, Wang Z, Tian Z, Ren L, et al. Toward High Carrier Mobility and Low Contact Resistance: Laser Cleaning of PMMA Residues on Graphene Surfaces. Nano-Micro Letters 2016, 8(4): 336-346.